\input harvmac
\input graphicx
\input color

\def\Title#1#2{\rightline{#1}\ifx\answ\bigans\nopagenumbers\pageno0\vskip1in
\else\pageno1\vskip.8in\fi \centerline{\titlefont #2}\vskip .5in}

%
%
\ifx\includegraphics\UnDeFiNeD\message{(NO graphicx.tex, FIGURES WILL BE IGNORED)}
\def\figin#1{\vskip2in}
\else\message{(FIGURES WILL BE INCLUDED)}\def\figin#1{#1}
\fi
\def\Fig#1{Fig.~\the\figno\xdef#1{Fig.~\the\figno}\global\advance\figno
 by1}
%
%
%
%
\def\Ifig#1#2#3#4{
\goodbreak\midinsert
\figin{\centerline{
\includegraphics[width=#4truein]{#3}}}
\narrower\narrower\noindent{\footnotefont
{\bf #1:}  #2\par}
\endinsert
}

\font\ticp=cmcsc10

\def \purge#1 {\textcolor{magenta}{#1}}
\def \new#1 {\textcolor{blue}{#1}}
\def\comment#1{}

\def\\{\cr}
\def\text#1{{\rm #1}}
\def\frac#1#2{{#1\over#2}}

\def\hf{{1\over 2}} 
\def\calo{{\cal O}}

\def\caln{{\cal N}}
\def\calm{{\cal M}}
\def\calh{{\cal H}}
\def\cala{{\cal A}}

\def\call{{\cal L}}
\def\calh{{\cal H}}

\def\roughly#1{\mathrel{\raise.3ex\hbox{$#1$\kern-.75em\lower1ex\hbox{$\sim$}}}}
\font\bbbi=msbm10 
\def\mathbb#1{\hbox{\bbbi #1}}

 \def\xb{{\bar x}}
 \def\ehat{{\hat e}}

\def\mthsu{\mathsurround=0pt  }
\def\leftrightarrowfill{$\mthsu \mathord\leftarrow\mkern-6mu\cleaders
  \hbox{$\mkern-2mu \mathord- \mkern-2mu$}\hfill
  \mkern-6mu\mathord\rightarrow$}
\def\overleftrightarrow#1{\vbox{\ialign{##\crcr\leftrightarrowfill\crcr\noalign{\kern-1pt\nointerlineskip}$\hfil\displaystyle{#1}\hfil$\crcr}}}
\overfullrule=0pt

%
%
\lref\MaldAdS{
  J.~M.~Maldacena,
  ``The Large N limit of superconformal field theories and supergravity,''
Int.\ J.\ Theor.\ Phys.\  {\bf 38}, 1113 (1999), [Adv.\ Theor.\ Math.\ Phys.\  {\bf 2}, 231 (1998)].
[hep-th/9711200].
}
\lref\MaroUnit{
  D.~Marolf,
  ``Unitarity and Holography in Gravitational Physics,''
Phys.\ Rev.\ D {\bf 79}, 044010 (2009).
[arXiv:0808.2842 [gr-qc]].
}
\lref\MaroHolo{
  D.~Marolf,
  ``Holographic Thought Experiments,''
Phys.\ Rev.\ D {\bf 79}, 024029 (2009).
[arXiv:0808.2845 [gr-qc]].
}
\lref\MaroNoST{
  D.~Marolf,
  ``Holography without strings?,''
Class.\ Quant.\ Grav.\  {\bf 31}, 015008 (2014).
[arXiv:1308.1977 [hep-th]].
}
\lref\QFG{
  S.~B.~Giddings,
  ``Quantum-first gravity,''
Found.\ Phys.\  {\bf 49}, no. 3, 177 (2019).
[arXiv:1803.04973 [hep-th]].
}
\lref\SSS{
  P.~Saad, S.~H.~Shenker and D.~Stanford,
  ``JT gravity as a matrix integral,''
[arXiv:1903.11115 [hep-th]].
}
\lref\Cole{
  S.~R.~Coleman,
  ``Black Holes as Red Herrings: Topological Fluctuations and the Loss of Quantum Coherence,''
Nucl.\ Phys.\ B {\bf 307}, 867 (1988)..
}
\lref\GiStInco{
  S.~B.~Giddings and A.~Strominger,
  ``Loss of Incoherence and Determination of Coupling Constants in Quantum Gravity,''
Nucl.\ Phys.\ B {\bf 307}, 854 (1988)..
}
\lref\GiStthird{
  S.~B.~Giddings and A.~Strominger,
  ``Baby Universes, Third Quantization and the Cosmological Constant,''
Nucl.\ Phys.\ B {\bf 321}, 481 (1989)..
}
\lref\MaMa{
  D.~Marolf and H.~Maxfield,
  ``Transcending the ensemble: baby universes, spacetime wormholes, and the order and disorder of black hole information,''
[arXiv:2002.08950 [hep-th]].
}
\lref\Locbdt{
  S.~B.~Giddings and M.~Lippert,
  ``The Information paradox and the locality bound,''
Phys.\ Rev.\ D {\bf 69}, 124019 (2004).
[hep-th/0402073].
}
\lref\JLMS{
  D.~L.~Jafferis, A.~Lewkowycz, J.~Maldacena and S.~J.~Suh,
  ``Relative entropy equals bulk relative entropy,''
JHEP {\bf 1606}, 004 (2016).
[arXiv:1512.06431 [hep-th]].
}
\lref\BCS{
  N.~Bao, S.~M.~Carroll and A.~Singh,
  ``The Hilbert Space of Quantum Gravity Is Locally Finite-Dimensional,''
Int.\ J.\ Mod.\ Phys.\ D {\bf 26}, no. 12, 1743013 (2017).
[arXiv:1704.00066 [hep-th]].
}
\lref\Chru{Piotr T. Chru\'sciel, ``Anti-gravity \`a la Carlotto-Schoen,"  arXiv:1611.01808 [math.DG].}
\lref\DoGithree{
  W.~Donnelly and S.~B.~Giddings,
  ``How is quantum information localized in gravity?,''
Phys.\ Rev.\ D {\bf 96}, no. 8, 086013 (2017).
[arXiv:1706.03104 [hep-th]].
}
\lref\DoGifour{
  W.~Donnelly and S.~B.~Giddings,
  ``Gravitational splitting at first order: Quantum information localization in gravity,''
Phys.\ Rev.\ D {\bf 98}, no. 8, 086006 (2018).
[arXiv:1805.11095 [hep-th]].
}
\lref\HKLL{
  A.~Hamilton, D.~N.~Kabat, G.~Lifschytz and D.~A.~Lowe,
  ``Holographic representation of local bulk operators,''
Phys.\ Rev.\ D {\bf 74}, 066009 (2006).
[hep-th/0606141].
}
\lref\FaLe{
  T.~Faulkner and A.~Lewkowycz,
  ``Bulk locality from modular flow,''
JHEP {\bf 1707}, 151 (2017).
[arXiv:1704.05464 [hep-th]].
}
\lref\QGQFA{
  S.~B.~Giddings,
  ``Quantum gravity: a quantum-first approach,''
LHEP {\bf 1}, no. 3, 1 (2018).
[arXiv:1805.06900 [hep-th]].
}
\lref\CHPSSW{
  J.~Cotler, P.~Hayden, G.~Penington, G.~Salton, B.~Swingle and M.~Walter,
  ``Entanglement Wedge Reconstruction via Universal Recovery Channels,''
Phys.\ Rev.\ X {\bf 9}, no. 3, 031011 (2019).
[arXiv:1704.05839 [hep-th]].
}
\lref\UQM{
  S.~B.~Giddings,
  ``Universal quantum mechanics,''
Phys.\ Rev.\ D {\bf 78}, 084004 (2008).
[arXiv:0711.0757 [quant-ph]].
}
\lref\Haag{R. Haag, {\sl Local quantum physics, fields, particles, algebras}, Springer (Berlin, 1996).}
\lref\SGalg{
  S.~B.~Giddings,
 ``Hilbert space structure in quantum gravity: an algebraic perspective,''
JHEP {\bf 1512}, 099 (2015).
[arXiv:1503.08207 [hep-th]].
}
\lref\locbdi{
  S.~B.~Giddings and M.~Lippert,
  ``Precursors, black holes, and a locality bound,''
Phys.\ Rev.\ D {\bf 65}, 024006 (2002).
[hep-th/0103231].
}
\lref\DoGione{
  W.~Donnelly and S.~B.~Giddings,
  ``Diffeomorphism-invariant observables and their nonlocal algebra,''
Phys.\ Rev.\ D {\bf 93}, no. 2, 024030 (2016), Erratum: [Phys.\ Rev.\ D {\bf 94}, no. 2, 029903 (2016)].
[arXiv:1507.07921 [hep-th]].
}
\lref\DoGitwo{
  W.~Donnelly and S.~B.~Giddings,
  ``Observables, gravitational dressing, and obstructions to locality and subsystems,''
Phys.\ Rev.\ D {\bf 94}, no. 10, 104038 (2016).
[arXiv:1607.01025 [hep-th]].
}
\lref\HMPS{
  I.~Heemskerk, D.~Marolf, J.~Polchinski and J.~Sully,
  ``Bulk and Transhorizon Measurements in AdS/CFT,''
JHEP {\bf 1210}, 165 (2012).
[arXiv:1201.3664 [hep-th]].
}
\lref\CCM{
  C.~Cao, S.~M.~Carroll and S.~Michalakis,
  ``Space from Hilbert Space: Recovering Geometry from Bulk Entanglement,''
Phys.\ Rev.\ D {\bf 95}, no. 2, 024031 (2017).
[arXiv:1606.08444 [hep-th]].
}
\lref\HeTe{
  M.~Henneaux and C.~Teitelboim,
  ``Asymptotically anti-De Sitter Spaces,''
Commun.\ Math.\ Phys.\  {\bf 98}, 391 (1985)..
}
\lref\CaSi{
  S.~M.~Carroll and A.~Singh,
  ``Mad-Dog Everettianism: Quantum Mechanics at Its Most Minimal,''
[arXiv:1801.08132 [quant-ph]].
}
\lref\Hartone{
  J.~B.~Hartle,
  ``The Quantum mechanics of cosmology,''
  in {\sl Quantum cosmology and baby universes : proceedings}, 7th Jerusalem Winter School for Theoretical Physics, Jerusalem, Israel, December 1989, ed. S. Coleman, J. Hartle, T. Piran, and S. Weinberg (World Scientific, 1991).
}
\lref\Harttwo{
  J.~B.~Hartle,
  ``Space-time coarse grainings in nonrelativistic quantum mechanics,''
  Phys.\ Rev.\  D {\bf 44}, 3173 (1991).
}
\lref\HartLH{
  J.~B.~Hartle,
  ``Space-Time Quantum Mechanics And The Quantum Mechanics Of Space-Time,''
  arXiv:gr-qc/9304006.
}
\lref\HartPuri{
  J.~B.~Hartle,
  ``Quantum Mechanics At The Planck Scale,''
  arXiv:gr-qc/9508023.
}
\lref\Eins{A. Einstein, ``Quanten-Mechanik und Wirklichkeit," Dialectica {\bf 2} (1948) 320.}
\lref\Howa{D. Howard, ``Einstein on locality and separability," Studies in History and Philosophy of Science Part A  {\bf 16} no. 3, (1985)
171.}
\lref\vanR{
  M.~Van Raamsdonk,
 ``Building up spacetime with quantum entanglement,''
Gen.\ Rel.\ Grav.\  {\bf 42}, 2323 (2010), [Int.\ J.\ Mod.\ Phys.\ D {\bf 19}, 2429 (2010)].
[arXiv:1005.3035 [hep-th]].
}
\lref\GiRo{
  S.~B.~Giddings and M.~Rota,
 ``Quantum information/entanglement transfer rates between subsystems,''
[arXiv:1710.00005 [quant-ph]].
}
\lref\BRSSZ{
  A.~R.~Brown, D.~A.~Roberts, L.~Susskind, B.~Swingle and Y.~Zhao,
  ``Complexity, action, and black holes,''
Phys.\ Rev.\ D {\bf 93}, no. 8, 086006 (2016).
[arXiv:1512.04993 [hep-th]].
}
\lref\Sussfall{
  L.~Susskind,
  ``Why do Things Fall?,''
[arXiv:1802.01198 [hep-th]].
}
\lref\NPGNL{
  S.~B.~Giddings,
  ``(Non)perturbative gravity, nonlocality, and nice slices,''
Phys.\ Rev.\ D {\bf 74}, 106009 (2006).
[hep-th/0606146].
}
\lref\ReTe{
  T.~Regge and C.~Teitelboim,
  ``Role of Surface Integrals in the Hamiltonian Formulation of General Relativity,''
Annals Phys.\  {\bf 88}, 286 (1974)..
}\lref\Heem{
  I.~Heemskerk,
  ``Construction of Bulk Fields with Gauge Redundancy,''
JHEP {\bf 1209}, 106 (2012).
[arXiv:1201.3666 [hep-th]].
}
\lref\KaLigrav{
  D.~Kabat and G.~Lifschytz,
  ``Decoding the hologram: Scalar fields interacting with gravity,''
Phys.\ Rev.\ D {\bf 89}, no. 6, 066010 (2014).
[arXiv:1311.3020 [hep-th]].
}
\lref\Zure{W. H. Zurek, ``Quantum darwinism, classical reality, and the 
randomness of quantum jumps," Physics Today {\bf 67} 44, 	arXiv:1412.5206.
}
\lref\LQGST{
  S.~B.~Giddings,
 ``Locality in quantum gravity and string theory,''
Phys.\ Rev.\ D {\bf 74}, 106006 (2006).
[hep-th/0604072].
}
\lref\GiKi{
  S.~B.~Giddings and A.~Kinsella,
  ``Gauge-invariant observables, gravitational dressings, and holography in AdS,''
JHEP {\bf 1811}, 074 (2018).
[arXiv:1802.01602 [hep-th]].
}
\lref\SGerice{
  S.~B.~Giddings,
  ``The gravitational S-matrix: Erice lectures,''
Subnucl.\ Ser.\  {\bf 48}, 93 (2013).
[arXiv:1105.2036 [hep-th]].
}
\lref\GKP{
  S.~S.~Gubser, I.~R.~Klebanov and A.~M.~Polyakov,
  ``Gauge theory correlators from noncritical string theory,''
Phys.\ Lett.\ B {\bf 428}, 105 (1998).
[hep-th/9802109].
}
\lref\Jacoholo{
  T.~Jacobson,
  ``Boundary unitarity and the black hole information paradox,''
Int.\ J.\ Mod.\ Phys.\ D {\bf 22}, 1342002 (2013).
[arXiv:1212.6944 [hep-th]].
}
\lref\DHW{
  X.~Dong, D.~Harlow and A.~C.~Wall,
  ``Reconstruction of Bulk Operators within the Entanglement Wedge in Gauge-Gravity Duality,''
Phys.\ Rev.\ Lett.\  {\bf 117}, no. 2, 021601 (2016).
[arXiv:1601.05416 [hep-th]].
}
\lref\JaNg{
  T.~Jacobson and P.~Nguyen,
  ``Diffeomorphism invariance and the black hole information paradox,''
Phys.\ Rev.\ D {\bf 100}, no. 4, 046002 (2019).
[arXiv:1904.04434 [gr-qc]].
}
\lref\Witt{
  E.~Witten,
  ``Anti-de Sitter space and holography,''
Adv.\ Theor.\ Math.\ Phys.\  {\bf 2}, 253 (1998).
[hep-th/9802150].
}
\lref\RyTa{
  S.~Ryu and T.~Takayanagi,
  ``Holographic derivation of entanglement entropy from AdS/CFT,''
Phys.\ Rev.\ Lett.\  {\bf 96}, 181602 (2006).
[hep-th/0603001].
}
\lref\BuVe{
  D.~Buchholz and R.~Verch,
  ``Scaling algebras and renormalization group in algebraic quantum field theory,''
Rev.\ Math.\ Phys.\  {\bf 7}, 1195 (1995).
[hep-th/9501063].
}
\lref\Yngv{
  J.~Yngvason,
  ``The Role of type III factors in quantum field theory,''
Rept.\ Math.\ Phys.\  {\bf 55}, 135 (2005).
[math-ph/0411058].
}
\lref\ZLL{
  P.~Zanardi, D.~A.~Lidar and S.~Lloyd,
 ``Quantum tensor product structures are observable induced,''
Phys.\ Rev.\ Lett.\  {\bf 92}, 060402 (2004).
[quant-ph/0308043].
}
\lref\Harl{
  D.~Harlow,
  ``Wormholes, Emergent Gauge Fields, and the Weak Gravity Conjecture,''
JHEP {\bf 1601}, 122 (2016).
[arXiv:1510.07911 [hep-th]].
}
\lref\GuJa{
  M.~Guica and D.~L.~Jafferis,
  ``On the construction of charged operators inside an eternal black hole,''
SciPost Phys.\  {\bf 3}, no. 2, 016 (2017).
[arXiv:1511.05627 [hep-th]].
}
\lref\CPR{
  J.~S.~Cotler, G.~R.~Penington and D.~H.~Ranard,
  ``Locality from the Spectrum,''
[arXiv:1702.06142 [quant-ph]].
}
\lref\PaRa{
  K.~Papadodimas and S.~Raju,
  ``Local Operators in the Eternal Black Hole,''
Phys.\ Rev.\ Lett.\  {\bf 115}, no. 21, 211601 (2015).
[arXiv:1502.06692 [hep-th]].
}
\lref\CoSc{
  J.~Corvino and R.~M.~Schoen,
  ``On the asymptotics for the vacuum Einstein constraint equations,''
J.\ Diff.\ Geom.\  {\bf 73}, no. 2, 185 (2006).
[gr-qc/0301071].
}
\lref\ChDe{
  P.~T.~Chrusciel and E.~Delay,
  ``On mapping properties of the general relativistic constraints operator in weighted function spaces, with applications,''
Mem.\ Soc.\ Math.\ France {\bf 94}, 1 (2003).
[gr-qc/0301073].
}
\lref\NVNLpost{
  S.~B.~Giddings,
  ``Nonviolent unitarization: basic postulates to soft quantum structure of black holes,''
JHEP {\bf 1712}, 047 (2017).
[arXiv:1701.08765 [hep-th]].
}
\lref\ADH{
  A.~Almheiri, X.~Dong and D.~Harlow,
  ``Bulk Locality and Quantum Error Correction in AdS/CFT,''
JHEP {\bf 1504}, 163 (2015).
[arXiv:1411.7041 [hep-th]].
}
\Title{
\vbox{\baselineskip12pt  
}}
{\vbox{\centerline{Holography and unitarity
} }}

\centerline{{\ticp 
Steven B. Giddings\footnote{$^\ast$}{Email address: giddings@ucsb.edu}
} }
\centerline{\sl Department of Physics}
\centerline{\sl University of California}
\centerline{\sl Santa Barbara, CA 93106}
\vskip.20in
\centerline{\bf Abstract}
If holography is an equivalence between quantum theories, one might expect it to be described by a map that is a bijective isometry between bulk and boundary Hilbert spaces, preserving the hamiltonian and symmetries.  Holography has been believed to be a property of gravitational (or string) theories, but not of non-gravitational theories; specifically Marolf has argued that it originates from the gauge symmetries and constraints of gravity.  These observations suggest study of the assumed holographic map as a function of the gravitational coupling $G$.  The zero coupling limit gives ordinary quantum field theory, and is therefore not necessarily expected to be holographic.  This, and the structure of gravity at non-zero $G$, raises important questions about the full map.  In particular, construction of  a holographic map appears to require as input a solution of the nonperturbative analog of the bulk gravitational constraints, that is, the unitary bulk evolution.  Moreover, examination of the candidate boundary algebra, including the boundary hamiltonian, reveals commutators that don't close in the usual fashion expected for a boundary theory.

\vskip.3in
\Date{}

\newsec{Introduction}

Holography, and in particular the AdS/CFT correspondence\refs{\MaldAdS}, has become a dominant theme in quantum gravity, but nonetheless its precise formulation and explanation remains controversial.  The purpose of this paper is to more carefully examine possible properties of a holographic correspondence, and in particular to investigate the role of gravitational effects.  

Gravity has been argued to play an essential role in holography\refs{\MaroUnit\MaroHolo\MaroNoST\Jacoholo-\JaNg}, via gauge invariance and the gravitational constraints.  This question can be examined systematically, beginning with scattering of states with weak gravitational fields, in large radius AdS.  Specifically, for such states, we expect to be able to study properties of the correspondence in an expansion in small Newton's constant $G$.  One interesting point of comparison, to infer the possible structure of holography, is the $G=0$ limit, which we expect to correspond to ordinary local quantum field theory (QFT) and not be holographic.  Then, the structure of gravitational effects can be investigated to higher orders in non-zero $G$.  Such study in fact suggests properties of the all-orders theory.

In addition to raising some puzzles, this analysis leads to two interesting conclusions.  The first is that in order to construct the candidate holographic map from Marolf's argument\refs{\MaroUnit\MaroHolo-\MaroNoST}, which relies on the equality between bulk and boundary hamiltonians, one apparently needs to solve the all-orders gravitational constraints, which is tantamount to knowing the bulk evolution.   Second, when one examines the candidate boundary algebra that follows from commutators of the boundary gravitational hamiltonian  with operators near the boundary, it does not exhibit the  closure properties that are expected for constructing a holographic map to a boundary theory.

This paper begins with some attention to setting up the basic framework; the next section proposes a definition of a holographic correspondence as a bijective isometry between the ``bulk" and ``boundary" Hilbert spaces, formalizing common beliefs, reviews other features of the correspondence, and raises some questions.  Section three outlines some expected properties of quantum gravity in AdS, particularly in the limit of weak gravitational fields.  Section four describes the $G=0$ QFT limit of the correspondence, and discusses properties that reinforce the expectation that this limit, with no gravity, is not holographic.  Finally, section five investigates the correspondence beginning with a perturbative expansion in $G$, but also inferring more general properties.  In particular, a key equation of holography, equality of the bulk and boundary hamiltonians, requires solution of the gravitational constraints, which in turn requires gravitational dressing of states and operators.  All-orders solution of the constraints appears to be needed to give the arguments for gravitational holography.  One can then inquire whether the resulting dressed operators have the desired properties; in particular, one finds that the candidate boundary algebra doesn't close in the fashion expected for a holographic description.  Similar issues are argued to arise in entanglement wedge reconstruction, which is briefly discussed.

\newsec{What is holography? (Formalizing beliefs, and some questions)}

We begin by stating some expected features of a holographic correspondence, such as that between gravity (or string theory) in $AdS_5$ and $\caln=4$ super-Yang Mills; this is largely a summary, but also sets up a proposed precise statement of the correspondence.

Specifically, the correspondence is commonly believed to be an equivalence between quantum theories.  The ``boundary" theory is an ordinary quantum field theory (QFT), with conformal invariance, and so should have a Hilbert space $\calh_\partial$ and operator algebra $\cala_\partial$ with standard QFT properties.   The ``bulk" theory is quantum gravity; if it is indeed a quantum theory, it is also expected to have a Hilbert space, $\calh_B$.  We also expect additional mathematical structure on $\calh_B$, {\it e.g.} an algebra of observables $\cala_B$, {\it etc.}, that is appropriate for describing gravity, although the precise form of this structure is still unknown.\foot{We thus approach quantum gravity from a ``quantum-first" perspective; for some recent discussion of progress on these questions, see \refs{\QFG}.}  

An equivalence between these quantum theories is  expected to be described by a map between bulk and boundary states
\eqn\hilbmap{\calh_B\, {\buildrel M\over\longrightarrow}\, \calh_\partial\ }
that is one-to-one, onto, and preserves inner products, that is, $M$ should be a {\it bijective isometry} (sometimes called a {\it global isometry}).  $M$ therefore also has an inverse map.  $M$ and its inverse  also induce a map between the operator algebras acting on the two Hilbert spaces; for example given an operator $\calo_B\in \cala_B$, the corresponding boundary operator is 
\eqn\opmap{\calo^\partial= \calm(\calo_B) =M \calo_B M^{-1}\ .}
Alternately, such an operator map $\calm$, together with an identification of the vacuum states, $|0\rangle_B\rightarrow |0\rangle_\partial$, can be thought of as defining a map $M$.

One also expects equivalent global symmetry group $SO(d,2)$ acting on the two Hilbert spaces, in the case of gravity in asymptotically $AdS_{d+1}$, and an equivalence between hamiltonians $H_B$ and $H^\partial$ via \opmap.    In the standard correspondence involving $\caln=4$ super-Yang Mills one of course expects the bulk spacetime to be $AdS_5\times S^5$, but we follow common practice of ignoring the $S^5$ as nontrivial dependence on these directions is expected to be handled by standard Kaluza-Klein methods.

The parameters of the bulk theory are the $D=d+1$-dimensional Newton's constant $G_D$ and the $AdS$ radius $R$ determined by the cosmological constant.  The parameters of a boundary $SU(N)$ CFT are the coupling $g_{YM}$, and $N$.  For the correspondence with $AdS_5\times S^5$ the parameters are related by
\eqn\AdsR{{R^4\over l_p^4} = N}
where $l_p\sim G_{10}^{1/8}$ is the Planck length.  If the bulk is a string theory, one also has the string length parameter $l_s=l_p/\sqrt{g_{YM}}$; we will primarily focus on gravitational effects, for which this is not relevant.

This discussion leaves us with a number of questions.  A first question is, what is the map $M$?  This is of course closely connected with a second question of describing $\calh_B$ and $\cala_B$. Specifically, as we will outline in the next section, we expect to have an approximate description of the bulk Hilbert space and algebra in a ``correspondence" limit of weak gravity.  But, the problem of quantum gravity can be viewed as that of finding the complete mathematical structure underlying bulk physics, which matches onto such a weak gravity description in this limit\refs{\QFG,\QGQFA}.  Obviously, if we {\it did} have a complete specification of $M$, that could furnish such structure.  (Conversely, if we understood $\calh_B$ and $\cala_B$, we might think of that as defining a boundary theory with Hilbert space $\calh_\partial$.)

A more recent thread in the study of holography considers whether {\it ensembles} of theories, with different values for the couplings, are relevant (see {\it e.g.} \refs{\SSS}).  The simplest possibility is that an ensemble of boundary theories corresponds to an ensemble of bulk theories, through an ensemble of maps such as \hilbmap.  However, it is also possible that a bulk theory with an enlarged Hilbert space, due to topology-changing processes, induces an effective ensemble of theories \refs{\Cole\GiStInco\GiStthird-\MaMa}, and there is significant current discussion about the exact role of such ensembles.  This paper will focus on the traditional view of holography, as a map between single theories, but is also expected to contain lessons for the case of ensembles.

\newsec{Expectations for bulk quantum gravity}

While we don't presently understand the complete mathematical structure on the bulk $\calh_B$, we do expect that we have a good approximate description of its structure in a weak gravity limit.  Specifically, assume that $R\gg l_p$, so $N\gg1$.  Then, for example, we expect that there are states with bulk energies $E\ll 1/l_p$ which behave just like QFT states on a background AdS spacetime, and in particular induce very small gravitational perturbations to the geometry;\foot{To avoid string scale effects, one also may require $R\gg l_s$, or $g_{YM}^2N\gg 1$, and a corresponding condition on energies.} moreover, there are such states with $E\gg 1/R$ which are well-localized compared to $R$.  Since some $SO(d,2)$ transformations act like boosts, this symmetry also implies that there are equivalent states that have superplanckian energies.  The space of such low-energy states, and their $SO(d,2)$ images, will be called $\calh_{LE}\subset \calh_B$.

There are additional states for which we expect a weak gravity description to be valid.  For example, the gravitational scattering of Oumuamua with our solar system certainly had center-of-mass (CM) energy $E_{CM}\gg 1/l_p$, yet is expected to be well described within weak gravity, and equally so if we lived in an AdS universe with $R\gg 10^{10} ly$.  Specifically, we expect a valid description of physics via weak gravity and local QFT to hold in scattering where CM energies of any subprocess (including multi-particle subprocesses) don't exceed a ``locality bound\refs{\locbdi\Locbdt-\LQGST}," given in terms of the CM separation between particles $\Delta x$
\eqn\locbdeq{E\roughly< {\Delta x^{D-3}\over G_D}\ }
(for multiparticle subprocesses, we take $\Delta x$ to be the bounding distance for the collection of particles).
For  energies violating the locality bound \locbdeq, we expect strong gravity effects to be important, and a description via local QFT likely fails.   Notice that this criterion also allows ultraplanckian high-energy collisions, at sufficiently large impact parameter.  The space of states respecting such a locality bound, and their $SO(d,2)$ images, will be called $\calh_{LB}\subset \calh_B$.  Elsewhere in the literature, the terminology\ADH\  ``code subspace" is sometimes used to describe a similarly-defined space of states.

In either $\calh_{LE}$ or $\calh_{LB}$, one expects that gravitational effects in amplitudes are accounted for via perturbative gravity,   incorporating graviton exchange and radiation; in the case of  $\calh_{LB}$, exchange of multiple gravitons (hence, loops) is relevant, but can be, {\it e.g.}, treated as eikonalized single graviton exchange, so that strong gravity effects are not relevant.  This can be explained in terms of the small momentum transfer carried by any single graviton line (due to ``momentum fractionation"); for further discussion, see, {\it e.g.}, \refs{\SGerice}.  This weak approximation fails when the locality bound \locbdeq\ is violated.

We moreover expect that there are subspaces of $\calh_{LE}$ or $\calh_{LB}$ in which gravity is completely negligible; one can consider states where $l_pE$ is negligible, or in the latter case, those in which collision energies yield negligible $G_D E/\Delta x^{D-3}$.  One can also formally say that these states arise from taking the limit $G_D\rightarrow0$, though the characterization in terms of energy is more accurate since $G_D$ is a dimensionful parameter.  Such states and their evolution can be thought of as being accurately described entirely within QFT.  

\newsec{What holography isn't}

Holography has been believed to be a property of theories that are gravitational, or string theories, and so it is conversely {\it not} an expected property of non-gravitational local QFTs, including those on an $AdS$ background.  This section will briefly explore properties of such theories that support the latter expectation.

Consider a QFT on a background spacetime; for concreteness, one could think of a scalar field theory with lagrangian $\call(\phi)$, or generalize to include other fields.  Such a theory is expected to define a Hilbert space $\calh_B$ with algebra of observables $\cala_B$.  In a local QFT, this algebra has additional structure, corresponding to the background spacetime.  Namely, in the algebraic approach\foot{See, for example, \refs{\Haag} for a review.} one studies the {\it net} of subalgebras of $\cala_B$ that are associated to open spatial regions, or to corresponding causal diamonds.  Via inclusion, intersection, {\it etc.} relationships, these ``mirror" the topological structure of the spacetime manifold.  They also capture the local/causal structure; locality is the statement that subalgebras associated to spacelike separated regions commute.

  \Ifig{\Fig\Adsdoms}{Shown is a small causal diamond $D_0$ near the center of AdS, and a causal domain $D_1$ near its boundary.}{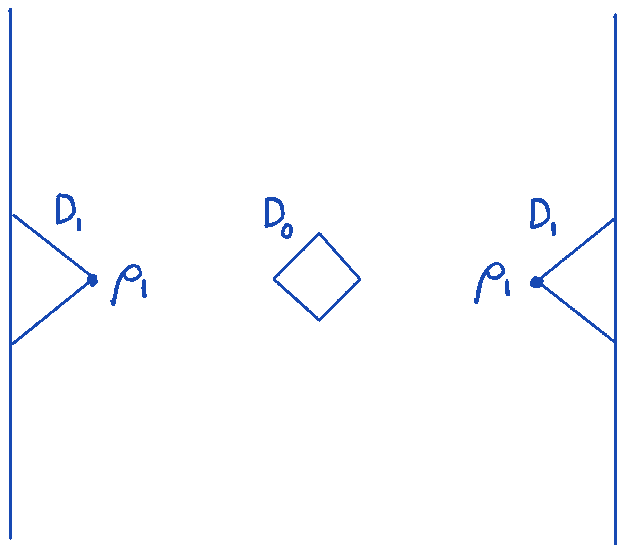}{2.25}

Consider specifically a background AdS spacetime.  It has been expected that in holographic theories, the operators of the boundary theory correspond to limits of bulk operators as they approach the boundary, as was described for example in \refs{\GKP,\Witt} (the ``extrapolate dictionary").\foot{The expectation that boundary operators arise from limits of bulk operators extends to nonlocal boundary operators; for example, in string theory, Wilson loops have been argued to correspond to operators that create corresponding long bulk strings.}  This is, however, at odds with the locality property of QFT.  Consider a small causal diamond $D_0$ near the center of AdS.  For example, suppose we work in the global coordinates where AdS takes the form
\eqn\adsmet{ds^2= {R^2\over \cos^2\rho}\left(-d\tau^2 + d\rho^2 + \sin^2\rho d\Omega_{D-2}^2\right)\ ,}
and let $D_0$ be centered on $\tau=\rho=0$.  Consider also the domain of dependence\foot{Given a closed spatial region this is the union of the future and past Cauchy development; for such an AdS region extending to the boundary, one also needs data on the boundary of AdS.}  $D_1$ of the annular region $\rho >\rho_1<\pi/2$, 
as in \Adsdoms; the union of diamonds contained in this domain corresponds to a subalgebra, which one therefore expects to contain a subalgebra equivalent to the boundary algebra $\cala_\partial$ via \opmap.  However, there are states in $\calh_B$ which are indistinguishable from $|0\rangle_B$ via such a subalgebra $\cala_\partial$.  Concretely, if one considers a bulk scalar field $\phi(x)$, and a c-number source $J(x)$ with support only in $D_0$, then the state
\eqn\Jstate{|J\rangle = e^{i \int d^Dx \sqrt{-g} J(x)\phi(x)} |0\rangle_B }
has the property that
\eqn\Jvev{\langle J| A' |J\rangle = {}_B\langle 0| A' |0\rangle_B}
for any operator $A'\in \cala_{D_1}$.  No operators (including composite nonlocal operators) defined from a boundary limit of bulk operators in the region $D_1$ register the nontrivial state $|J\rangle$. Put differently, the operator $\exp\{i\int J\phi\}$ in \Jstate\ doesn't obey the time slice axiom with respect to the hypothesized boundary algebra -- it commutes with boundary operators, but is nontrivial.

In other words, locality implies that there is bulk information that is not present in any such ``boundary theory." One expects to also be able to explain this by construction of a split vacuum\foot{See \Haag, and references therein; we assume such constructions extend to nontrivial spacetimes such as AdS.}, which is a bulk state $|U_{D_0}\rangle$ such that
\eqn\spltdef{\langle U_{D_0}| AA'|U_{D_0}\rangle = {}_B\langle 0| A |0\rangle_B\ {}_B\langle 0|A' |0\rangle_B}
for any $A\in \cala_{D_0}$ and $A'\in \cala_{D_1}$; thus any two states $A_1|U_{D_0}\rangle$, $A_2|U_{D_0}\rangle$, with $A_1,A_2\in \cala_{D_0}$,  are also indistinguishable in the boundary algebra.

While these statements are expected to capture the essential contrast between holography and locality, it is also worth examining other kinds of statements made about holographic theories, in this non-holographic context.

One approach to holography is to consider evolution to the boundary.  For example, once the causal future of $D_0$ intersects the AdS boundary, one expects that it is possible for near-boundary operators to distinguish the state $|J\rangle$ of \Jstate\ from the vacuum.  Then, if there are corresponding boundary operators in a boundary theory, one might also na\"\i vely expect that the boundary evolution could be used to evolve the pertinent boundary operators back in time to construct boundary operators at time $\tau=0$ that register the state $|J\rangle$, in contrast to the preceding discussion.

However, this fails in QFT because the bulk hamiltonian $H_B$ does not map to a boundary hamiltonian $H^\partial\in \cala_\partial$ that preserves the boundary algebra.  
A simplified version of this, capturing the essential point, can be illustrated   by considering a free scalar hamiltonian in flat 3d space, 
\eqn\flatH{H_B=\hf\int d^3x  [\pi^2 + (\nabla \phi)^2]\ ,}
and, say, a boundary given by the plane $z=0$.  Here, we do not expect a holographic description of the theory at $z<0$.  Indeed, suppose that $\phi(t,x,y,0)$ is in a boundary algebra.  Then $[H_B,\phi(t,x,y,0)]= -i\pi(t,x,y,0)$, and so $[H_B,[H_B,\phi(t,x,y,0)]] = - \sum_i\partial_i^2\phi(t,x,y,0)$.  While $x$ and $y$ derivatives of $\phi$ of course lie in the boundary algebra, iteration shows that commutators with $H$ do not close on any finite number of $z$ derivatives at the boundary.  This reflects the fact that information can reach $z=0$ from $z<0$.  Similar statements hold for QFT in an AdS background.

Of course a simple contrast to this is  a situation where there is a bona-fide (2+1)-d hamiltonian, for example
\eqn\tdH{H=\hf\int d^2x  [\pi^2 + (\partial_x \phi)^2+ (\partial_y \phi)^2]\ .}
Then $[H,\phi(t,x,y)] = -i\pi(t,x,y)$ and $[H,[H,\phi(t,x,y)]] = -\partial_x^2\phi(t,x,y)-\partial_y^2\phi(t,x,y)$, and so the algebra closes -- commutators are determined in terms of elements of the algebra.  This is just a different way of saying that  (2+1)-d evolution with hamiltonian \tdH\ is a well-defined Cauchy problem. We would expect a general boundary theory to behave this way, and this provides a criterion for closure of the boundary algebra.

  \Ifig{\Fig\Adsregs}{A boundary region $A_\partial$, and its corresponding Ryu-Takayanagi surface $\chi(A_\partial)$ and bulk domain $A_B$.}{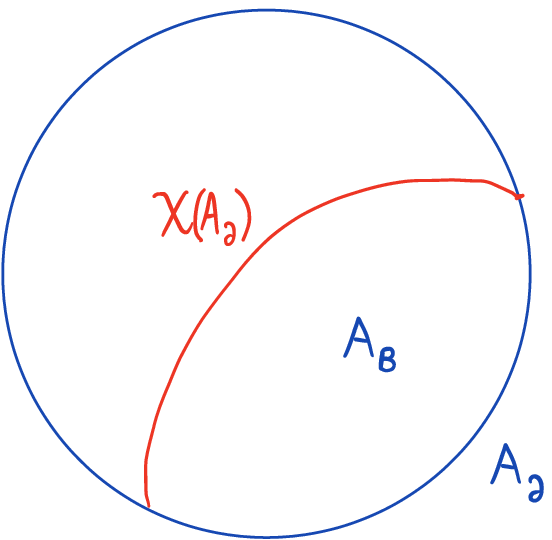}{1.75}

Another approach to holography is that of entanglement wedge reconstruction (EWR)\refs{\DHW\CHPSSW-\FaLe}, and so an obvious question is what this might say in the case of a non-holographic theory.  As discussed in \DHW, EWR is based on the key equation relating entropies,
\eqn\entropeq{S(\rho_{A_\partial}) = S(\rho_{A_B}) +{1\over 4G_D}  {\rm Tr} [\rho_{A_B}{\rm Area}(\chi_{A_\partial}) +\cdots]\ ,}
where $A_\partial$ is a boundary spatial region, and $A_B$ is the bulk spatial region bounded by $A_\partial$ and the Ryu-Takayanagi surface\refs{\RyTa} $\chi(A_\partial)$ of $A_\partial$; see \Adsregs.  The density matrices are supposed to be calculated by tracing out degrees of freedom in the complementary regions, in some fiducial state.  A first question is what \entropeq\ states in the QFT limit of $G_D\rightarrow0$.  If $S(\rho_{A_B})$ is defined with a distance cutoff $\epsilon\gg l_p$, the second term on the right dominates, and gives an infinite entropy for $\rho_{A_\partial}$.  Then the bulk QFT entropy of $A_B$ is a small, relative $\calo(G_D)\rightarrow0$ correction.  So, we don't seem to find a particularly useful statement at $G_D=0$.  We  return to the question of the nonzero $G_D$ version of EWR below.

\newsec{Holography at nonzero $G_D$?}

Already the preceding discussion raises a puzzle.  If the holographic map $M$ of \hilbmap\ exists for arbitrary $G_D$, we might expect it to exist in the limit $G_D\rightarrow0$.  But we have just seen that such a map is not expected in this limit, unless the behavior changes qualitatively at $G_D=0$.  
More precisely, for any $G_D$, and for sufficiently large $R$, we expect to be able to define the low energy subspaces where gravity is irrelevant and all bulk physics is described by local QFT, as described in section 3.  The holographic map $M$ of \hilbmap\ must act on these subspaces, yet we have no candidate definition of a holographic boundary theory dual to the resulting QFT.

In order to investigate this puzzle more closely, as well as expand on the discussion of EWR, this section will begin by considering the case of ``small but nonzero $G_D$," which we have seen is really  a condition specifying a subsector of states of the $R/l_p\gg 1$ theory.  In fact, Marolf\refs{\MaroUnit\MaroHolo-\MaroNoST} (for further discussion, see also \refs{\Jacoholo,\JaNg}) has argued that holography arises from specific features unique to gravity; we can explore the validity of this proposal, beginning with the small $G_D$ limit.

Marolf's proposed explanation of holography can be thought of as a gravitational upgrade of the argument involving propagation to the boundary described in the preceding section.  Schematically, we argued that a state like $|J\rangle$ of \Jstate\ could be registered by boundary operators in the causal future of the support of the source $J$.  However, in gravity, the hamiltonian is known to be a surface term, and thus is expected to lie in the boundary algebra.  This suggests that the boundary operators that register $|J\rangle$ can now be evolved back to boundary operators at $\tau=0$, giving operators that can detect the nontrivial structure of the state outside the causal future of the source.

\subsec{Bulk and boundary hamiltonians}

It is important to examine this argument more closely.  First, the general relation between the bulk and boundary hamiltonians in AdS is given by\ReTe\ 
\eqn\hamilreln{H_B = H_\partial + \int_\Sigma d\Sigma\ C_\xi }
(for an explicit derivation, see \GiKi).
Here $\Sigma$ is a constant-$\tau$ slice, and $d\Sigma$ and $n^\mu$  are its volume element and unit normal.  $C_\xi$ is a projection of the constraints 
\eqn\constraints{ C_\mu =  \left(T_{\mu\nu} - {\Lambda\over 8\pi G_D} g_{\mu\nu} -  {1\over 8\pi G_D}G_{\mu\nu}\right)n^\nu  }
into $\xi=(1,\vec 0)$, $C_\xi = C_\mu \xi^\mu$. Here $T_{\mu\nu}$ is the matter stress tensor, $G_{\mu\nu}$ the Einstein tensor and $\Lambda$ is the cosmological constant.  Note that here $H_\partial$ is a boundary limit of a bulk operator; the question of its map to $\cala_\partial$ will be considered below.  The bulk and boundary hamiltonians are  equal only on states annihilated\foot{In a perturbative quantization, it appears that the correct statement is that they be weakly annihilated \refs{\DoGifour}.} by the constraint $C_\xi$; for operators, the equality of their commutators with $H_B$ or $H_\partial$ follows if the operators  commute with the constraints.\foot{Eq.~\hamilreln\ involves only the temporal constraint $C_\tau=0$, but $SO(d,2)$ symmetry relates different time evolutions, and so in general one must also solve the spatial constraints $C_i=0$.}  Since the constraints generate gauge transformations (diffeomorphisms that vanish at infinity), this condition specifies gauge-invariant operators; examples of these are gravitationally dressed versions of operators of the underlying QFT that is coupled to gravity.

\subsec{Dressed operators}

Specifically, in a theory describing some matter coupled to gravity, we expect the constraints \constraints\ to follow from the more complete theory, in the weak-gravity regime appropriate to $\calh_{LE}$ or $\calh_{LB}$.  One can work perturbatively in $\kappa^2 = 32\pi G_D$, and construct dressed operators.  In particular, to leading order, explicit expressions have been found in \refs{\DoGione,\GiKi,\QGQFA} (for earlier related discussion see \refs{\Heem,\KaLigrav}).    Considering the case of a scalar $\phi(x)$ coupled to gravity, a dressed version takes the form
\eqn\phidress{\Phi(\xb) = \phi(X_g(\xb))\ .}
Here $X^\mu_g(\xb)$ are functionals of the metric, which also depend on gauge-invariant parameters $\xb$ that specify the location of the operator.  
Under a diffeomorphism $f(x)$, $\phi\rightarrow \phi(f^{-1}(x))$; for  an infinitesimal diffeomorphism $f(x)=x+\kappa \xi(x)$, 
\eqn\diffphi{\delta_{\kappa\xi} \phi = -\kappa\xi^\mu\partial_\mu\phi\ .}
In order for $\Phi(\xb)$ to be gauge invariant, we need the transformation $X_g(\xb)\rightarrow f(X_g(\xb))$, or infinitesimally,
\eqn\diffX{\delta_{\kappa\xi} X_g(\xb) = \kappa \xi^\mu(\xb)\ .}

For example, such dressings, which we write as  $X_g(\xb)= \xb+V(\xb)$, may be explicitly constructed for metrics that are perturbations about flat space,
\eqn\flatpert{g_{\mu\nu}=g^0_{\mu\nu}+\kappa h_{\mu\nu}\ ,}
with $g^0=\eta$ (the power of $\kappa$ is introduced so that $h$ and its commutators are canonically normalized).
An example is the gravitational line dressing of \refs{\DoGione,\QGQFA}, which for a general curve $\Gamma$ connecting $\xb$ to infinity, takes the form  
\eqn\linedress{V_\Gamma^\mu(\xb) = {\kappa\over 2} \int_\xb^\infty dx^{\nu} \left\{ h_{\mu\nu}(x) + \int_{x}^\infty dx^{\prime\lambda}\left[\partial_\mu h_{\nu\lambda}(x') - \partial_\nu h_{\mu\lambda}(x')\right]\right\}+\calo(\kappa^2)\ .}
Using the transformation for $h$,
\eqn\htrans{\delta_{\kappa\xi} h_{\mu\nu} = -\partial_\mu\xi_\nu -\partial_\nu\xi_\mu\ ,}
$V^\mu_\Gamma$ is easily seen to satisfy \diffX, and, when $\Phi$ acts on a state to create a $\phi$ particle, creates a corresponding line-like gravitational field.  Another such dressing\DoGione\ is found by taking an angular average of $V^\mu_\Gamma$ over straight lines to infinity, and creates a ``Coulomb" configuration that behaves like a linearized Schwarzschild field of a particle created by $\phi(x)$.  There are an infinity of such perturbative dressings, which create gravitational fields differing by radiation (sourceless) gravitational fields. 

Analogous  expressions for dressings perturbing about AdS are explicitly constructed in \GiKi.  There, for a simple form of the line-dressing, the curve $\Gamma$ should be a geodesic to infinity.  Averaging over such curves gives Coulomb or more general dressings.  The resulting dressed operators $\Phi_V(\xb)=\phi(\xb+V(\xb))$
 then commute with the constraints, implying 
\eqn\hcomm{[H_B,\Phi_V(\xb)] = [H_\partial,\Phi_V(\xb)]\ .}

\subsec{Exploring gravitational holography}

With this preparation, we can explicitly study  the arguments for gravitational holography\refs{\MaroUnit\MaroHolo\MaroNoST\Jacoholo-\JaNg}.  When we do so, various questions are encountered.\foot{For earlier discussion of some of these, see \GiKi.}

A first question regards the structure of the extrapolate map, namely the form of the map $\calm(\Phi_V(\xb))$ of \opmap, as $\xb$ approaches the boundary.  For a given underlying QFT operator $\phi(x)$, there are many (infinitely many, at the perturbative level) corresponding dressed operators $\Phi_V(\xb)$, which
create inequivalent physical states.  In general these should correspond to different boundary operators, differing at order $G_D$, or in $N^{-2}$ in the case of $AdS_5$.  For example, the dressing $V$ could connect to the boundary point that $\xb$ limits to, or to a different one.  The details of such a correspondence remain to be worked out.  But, suppose we grant that in general there is a map \opmap\ between such dressed operators as they approach the boundary, collectively denoted $\calo_i(b)$, and a collection of possibly nonlocal boundary operators $\calo^\partial_i(b)\in\cala_\partial$.  Here boundary points $b^\alpha=(\tau,\ehat)$ are parameterized using a unit vector $\ehat$ on  $S^{D-2}$.

If such a map is assumed, the next question regards how to invert it to determine $\Phi_V(\xb)$ for an $\xb$ that locates the operator near the center of AdS, in terms of boundary operators.  The Heisenberg picture operators evolve according to
\eqn\heisevol{\Phi_V(\tau,x^i) = e^{i H_B \tau}\Phi_V(0,x^i) e^{-i H_B \tau}\ ;}
indeed, as an alternative to \hamilreln, $H_B$ can be equivalently written in the usual form
\eqn\bulkH{H_B = \int_\Sigma d\Sigma n^\tau (T_{\tau\tau} + t_{\tau\tau})}
where $ t_{\tau\tau}$ is an effective stress tensor for the gravitational field.  The evolution \heisevol\ implies Heisenberg equations
\eqn\heiseqn{\dot \Phi_V = i [H_B,\Phi_V]\quad ,\quad \ddot \Phi_V =-[H_B, [H_B,\Phi_V]]\ .}
Suppose, now, that we can find appropriate Green functions that solve these and express $\Phi_V(\xb)$, for $\xb\approx 0$, in terms of the boundary limits $\calo_i(b)$ of the bulk operators.  At the linear level, such maps were proposed by \refs{\HKLL}, and a proposed nonlinear generalization\refs{\HMPS} takes the form
\eqn\nonlinG{\Phi_V(\xb) = \int db K_i(\xb,b)\calo_i(b) + \int db_1 db_2 K_{ij}(\xb,b_1,b_2)\calo_i(b_1)\calo_j(b_2) + \cdots}
(with summation convention used).
There are various questions about the actual construction of such expressions, but suppose we assume the existence of such expressions (and defer these questions for separate work).

We stress that the operators $\calo_i$ are boundary limits of bulk operators, {\it i.e.}, $\calo_i\in \cala_B$ (up to possible normalization issues).  
Thus, these, too,  evolve according to
\eqn\Bevol{\calo_i(\tau,\ehat) =e^{i H_B \tau}\calo_i(0,\ehat) e^{-i H_B \tau}\ .}
At this point, the structure of the expressions, with $H_B$ given by \bulkH, is the same as the (non-holographic) case of QFT, described in section 4.  What is supposed to make it holographic, according to Marolf\refs{\MaroUnit\MaroHolo-\MaroNoST} is the equality $H_B=H_\partial$, and the identification of the boundary limits with a boundary algebra.  We consider these points in turn.

\subsec{Holography and unitary evolution}

First, from \hamilreln, we recall that the equality  $H_B=H_\partial$ requires solving the constraints.
So, the constraints must be satisfied for any states for which we would like to have a holographic description -- or for the operators creating such states.\foot{Since the required time evolution in \nonlinG\ is over $\Delta\tau \sim 1$ (corresponding to a proper time $\sim R$), we require $H_B-H_\partial\ll 1$ for an accurate match.}  Consider, for example, a state with two particles scattering at superplanckian CM energy.  Initially , at $\tau=0$, these particles may have separations $\Delta x$ respecting the locality bound \locbdeq, and thus be expected to have a good weak gravity description.  But, suppose they then collide at an impact parameter violating the locality bound, and  make a quantum black hole.  Furthermore, suppose that this black hole then evaporates, with the formation and evaporation taking place on a time scale $\ll R$.  A holographic description of this process then appears to require solution of the full nonperturbative generalization of the constraints $C_\mu=0$.  

  \Ifig{\Fig\WDfig}{Shown is a ``Wheeler-DeWitt" patch associated with a given boundary time.  Since evolution between the slices $\Sigma_1$ and $\Sigma_2$ occurs at fixed boundary time, it should be generated by the constraints.  This evolution could include non-perturbative processes, such as black hole formation and evaporation.}{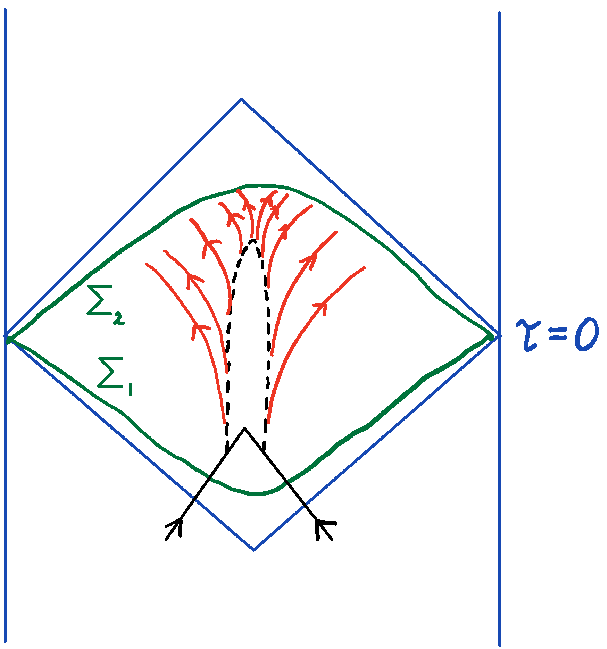}{2.2}

This is tantamount to having a description of the bulk unitary evolution.  This can be illustrated as in \WDfig, which shows a ``Wheeler-DeWitt" patch corresponding to a fixed boundary time, {\it e.g.} $\tau=0$.  We can associate to this time either the spacelike slice $\Sigma_1$ or $\Sigma_2$, and a single holographic description of the state would indicate that the states on these slices are equivalent.  Indeed, evolution between the slices is generated by the constraints.  For the process of black hole formation and decay, this must unitarily relate the incoming particles to the outgoing radiation, as indicated.

The conclusion of this discussion is that one needs to first have  a description of unitary bulk evolution, as input {\it in order to construct the gravitational holographic map}.  There is apparently  not such a holographic map that exists independent of unitary bulk evolution, so that unitary bulk evolution can be derived purely from the boundary evolution.

To see why this is important, suppose that we want to use the holographic correspondence to provide a unitary description of black hole evolution.  If there is a boundary theory we expect to have a description of unitary evolution in that theory, so if we are given the holographic map, that would determine the bulk evolution.  However, the preceding argument tells us that finding the holographic map requires solving the problem we hoped holography plus boundary evolution would solve for us, namely first solving the problem of unitary bulk evolution.

\subsec{A boundary algebra?}

The next question is that of the boundary algebra.  We have assumed that limits $\calo_i$ of bulk operators can be identified with boundary operators $\calo_i^\partial$ via the holographic map.  An important question is how this works for $H_\partial$, which we recall is in $\cala_B$, and the role this plays in commutators.  We first note that an explicit expression for $H_\partial$ in terms of the metric perturbation $h_{\mu\nu}$ was derived in \GiKi, using covariant canonical methods: 
\eqn\Hdel{\eqalign{H_\partial =\frac{2R^{D-4}}{\kappa} \oint d\Omega\,    \xi^\tau & \lim_{\rho\to \pi/2} (\tan\rho)^{D-3} \Biggl[\cot\rho\, \hat \nabla^a h_{a\rho} \cr &+(D-2) h_{\rho\rho} +\left(\frac{2-\cos^2\rho}{1-\cos^2\rho}+{\partial\over \partial\ln\cos\rho }\right) \hat g^{ab} h_{ab}\Biggr]\ ,}}
where $a,b$ denote components along the $S^{D-2}$, and $\hat g_{ab}$ and $\hat\nabla^a$ are its metric and covariant derivative. 
For metric perturbations with the na\"\i ve asymptotic behavior\refs{\HeTe}
\eqn\metasy{h_{\alpha\beta}\rightarrow \cos^{D-3}\rho\, {\bar h}_{\alpha\beta}(b^\gamma)
\quad,\quad h_{\rho\rho}\rightarrow \cos^{D-3}\rho\, {\bar h}_{\rho\rho}(b^\alpha)
 \quad,\quad h_{\rho\alpha}\rightarrow \cos^{D-2}\rho\, {\bar h}_{\rho\alpha}(b^\beta)\ ,}
this simplifies to
\eqn\Hsimp{ H_\partial= \frac{2}{\kappa} \oint d\Omega   \xi^\tau\ R^{D-4} [(D-2){\bar h}_{\rho\rho} +(D-1)\hat g^{ab} {\bar h}_{ab}]\  .}
However, dressings such as the AdS generalization of \linedress, and its averaged Coulomb form, create metric perturbations with slower falloff than \metasy; nonetheless, these have the correct commutators with the more general expression \Hdel, as shown in \GiKi.

Note that a common view is that one can work in an approximation where the backreaction of bulk operators on the metric is neglected -- that is, dressing is neglected.  However, the preceding discussion shows that this leads to violation of the equation identifying the bulk and boundary hamiltonians, $H_B=H_\partial$.  For example, neglecting the dressing reduces $\Phi_V$ to the operator $\phi(x)$, and the canonical commutators yield
\eqn\phicomm{[H_\partial,\phi(x)]=0\neq[H_B,\phi(x)]\ .}

While the complete dressing of an operator $\Phi_V$ may require solution of the nonperturbative constraints, the asymptotic form of $V$ near the AdS boundary is expected to be dominated by the leading perturbative structure.  One can investigate the commutator of $H_\partial$ with the AdS analog of the line $V$ of \linedress, or with an averaged version, and using canonical commutators for $h_{\mu\nu}$, one finds\GiKi
\eqn\HVcomm{[H_\partial,V^\mu(\xb)] = -i\delta^\mu_\tau +\calo(\kappa)\ }
the $\calo(\kappa)$ term comes from higher-order contributions to the dressing, and for consistency should describe time translation of the argument of $V$\refs{\DoGithree}.  
Then, the explicit commutators of $H_\partial$ with the dressed operators $\Phi_V(\xb)=\phi(\xb+V(\xb))$ yield
\eqn\HPhicomm{[H_\partial ,\Phi_V(\xb)] = -i\partial_{\bar \tau}\Phi_V(\xb)\quad , \quad [H_\partial ,[H_\partial ,\Phi_V(\xb)] ]=-\partial_{\bar \tau}^2 \Phi_V(\xb)\quad ,\quad {\it etc.}}

In the boundary limit $\xb\rightarrow \partial$, and using the map $\calm$ of \opmap, the commutators \HPhicomm\ should map to corresponding commutators of operators in the boundary algebra $\cala_\partial$.  But, the form of these commutators can be contrasted with what is expected for commutators with the hamiltonian in a boundary theory.  
For example, in the simple case of a scalar boundary theory, with hamiltonian \tdH, the commutators of the hamiltonian with the boundary operators gave equations of motion that determined the $t>0$ operators in terms of operators at $t=0$.  Specifically, these gave Cauchy evolution where the $t>0$ operators are determined by $\phi$ and $\partial_t\phi$ at $t=0$.  Likewise, if the boundary theory is a Yang Mills theory, we expect the same behavior.  Specifically, one can consider either gauge invariant operators, or consider the Yang-Mills fields and momenta in a specific gauge.  The time evolution of such operators is determined by a non-trivial equation
\eqn\Heiseom{\partial_\tau\calo^\partial = i [H^\partial,\calo^\partial]\ ,}
which, together with its iterations, determines all time derivatives in terms of $\tau=0$ operators with one or zero time derivatives.  That is, Yang Mills theory also has a standard canonical structure and Cauchy evolution.

However, using the commutators \HPhicomm, the Heisenberg equations \Heiseom\ imply
\eqn\triveq{\partial_\tau\calo^\partial=\partial_\tau\calo^\partial\ :}
that is, correct but {\it trivial} equations which provide no information and do not determine subsequent evolution of the operators in the absence of other information.  In this sense, the commutators \HPhicomm\ do not match boundary commutators that close in the expected fashion for a boundary theory.

There is an apparent logic to the failure of the boundary commutators arising from \HPhicomm\ to give an evolution equation that is independent of the bulk evolution.   The time evolution of operators has already been determined by solving the constraints, as described in the preceding subsection.  If there were a separate equation for boundary evolution, there is no a priori reason for the evolutions to agree.  The gravitational theory can escape this potential conflict by giving the trivial equations \triveq.

\subsec{Entanglement wedge reconstruction}

As noted above, a related approach to defining holography is that of entanglement wedge reconstruction\refs{\DHW\CHPSSW-\FaLe}.  A simple example of this considers a boundary spatial region $A_\partial$, that defines a Rindler wedge $D(A_B)$ with spatial base $A_B$ in the bulk.  (See \Adsregs.)
A bulk operator in $D(A_B)$ is written in the form of \nonlinG, where the boundary integrals are over $D(A_\partial)$ (see, {\it e.g.}, \refs{\FaLe}).
Then, once again the relations $H_B=H_\partial$ and between bulk and boundary algebras are argued to give an equal-time representation of the operator, in terms of boundary operators on $A_\partial$.  This argument thus relies on the same assumptions, namely bulk unitary evolution and appropriate closure of a boundary algebra, that have been critically examined above.

Specifically, the general arguments of \refs{\DHW} for EWR rely on the relation \entropeq.  This can likewise be seen to rest on solving the constraints.  Specifically, \refs{\JLMS} arrive at \entropeq\ from an underlying relation between modular hamiltonians,
\eqn\modham{K_\partial = {A\over 4G_D} + K_B\ ,}
where this expression is supposed to generalize to more general boundary regions $A_\partial$, and the area $A$ is that of the Ryu-Takayanagi surface\refs{\RyTa}.  In the Rindler case, the modular hamiltonians can be written in the form $K=\int \xi^\mu T_{\mu\nu} n^\nu$, where now $\xi^\mu$ is the vector generating the Rindler flow.  The equality of the modular hamiltonians in \modham\ thus arises, in analogy to \hamilreln, from the vanishing of the bulk constraints (Einstein's equations are indeed used in deriving \modham\ in sec.~4 of \JLMS).  This, and the analogous issue of the algebraic structure, suggests that EWR does not evade the arguments discussed in the preceding subsection.

\bigskip\bigskip\centerline{{\bf Acknowledgments}}\nobreak

This material is based upon work supported in part by the U.S. Department of Energy, Office of Science, under Award Number {DE-SC}0011702.  I thank  J. Maldacena for useful conversations, and T. Faulkner, T. Jacobson and D. Marolf for useful conversations and for helpful comments on a draft of this paper.

\listrefs
\end